\documentstyle[12pt]{article}

\newcommand{\be}{\begin{equation}}
\newcommand{\ee}{\end{equation}}
\newcommand{\bea}{\begin{eqnarray}}
\newcommand{\eea}{\end{eqnarray}}

\newcommand{\afin}{0.108(5)}
\newcommand{\bfin}{0.327(4)}
\newcommand{\gfin}{1.237(4)}
\newcommand{\dfin}{4.77(5)}
\newcommand{\pfin}{1.94(5)}

\newcommand{\aopt}{0.1092(40)}
\newcommand{\bopt}{0.3267(28)}
\newcommand{\gopt}{1.2373(23)}
\newcommand{\dopt}{4.791(39)}
\newcommand{\popt}{1.945(25)}

\title{Accurate estimates of 3D Ising critical exponents \\
       using the Coherent-Anomaly Method}
\author{Miroslav Kolesik\thanks{
        Permanent address:
        Institute of Physics, SAS, D\' ubravsk\' a cesta 9,
         Bratislava 842 28, Slovakia
        }\hspace {10pt}  and \hspace {10pt}
        Masuo Suzuki  \\
        Department of Physics, University of Tokyo, \\
          Bunkyo-ku, Tokyo 113, Japan}
\begin{document}
\maketitle

\centerline{cond-mat/9411109}

\begin{abstract}
An analysis of the critical behavior of the three-dimensional Ising
model using the coherent-anomaly method (CAM) is presented.  Various
sources of errors in CAM estimates of critical exponents are discussed,
and an improved scheme for the CAM data analysis is tested.  Using a
set of mean-field type approximations based on the variational series
expansion approach, accuracy comparable to the most precise
conventional methods has been achieved.  Our results for the
critical exponents are given by $\alpha=\afin$, $\beta=\bfin$,
$\gamma=\gfin$ and $\delta=\dfin$.


\bigskip
\noindent
{\bf Running title:}  3D Ising critical exponents using CAM
\end{abstract}

\section{Introduction}

This article is devoted to a study of critical properties of
the 3D Ising model by the coherent-anomaly method (CAM)
\cite{Suzuki86, Suzuki87, Katori}.  The Ising model has been
investigated intensively by various techniques, including the
CAM, for many years.  The long-standing interest in this model
is caused
by the fact that it is an ideal system for testing new methods
of studying critical phenomena as well as by its great
theoretical importance itself.

A great deal of effort made for understanding of this model has
been aimed at the calculation of critical exponents.  However,
there are still some uncertainty and scatter in the results
available in the literature on critical phenomena.  The most
accurate estimates are usually provided by extensive Monte Carlo
simulations, series expansions and field theoretic methods.  The
CAM, while proving its versatility and very good accuracy
in many physical systems (see e.g.  \cite{Suzuki94} for a recent
review), seems to exhibit slightly less precise results for 3D-Ising
critical exponents.

The purpose of the present work is twofold.  Firstly, we intend
to demonstrate that within the coherent-anomaly approach it is
possible to calculate critical indices with accuracy comparable
to the one reached by the most precise conventional methods.
Secondly, we want to discuss various sources of errors of the CAM
estimates in some detail in order to show a possible way for
further improvement of the results.

The article is organized as follows.  We give a brief summary of
the coherent-anomaly method in the next section for
convenience.  Then, we discuss the main sources of errors in the CAM
estimates of critical exponents from a general point of view, and
propose a new scheme to improve CAM results.  In the third
section we describe the generation of the series of mean-field
type approximations used in this study, which is based on the
variational series expansion (VSE).  Analysis and our results
together with a brief review of the available data for critical
exponents are encompassed in the fourth section.  Finally, we
present some concluding remarks in Section 5.

\section{Coherent-anomaly method}
\subsection{Brief summary of the CAM}

The coherent-anomaly method is a general approach for extracting the
true critical behavior of the system under investigation from the
systematic behavior of the classical criticality of a series of
mean-field type approximations \cite{Suzuki86, Suzuki87, Katori}.  It
is based on the coherent-anomaly scaling exhibited by certain
quantities in such a series.  The starting point for each CAM
analysis is thus a set of approximation for the given model.  For
each member of this set ( labeled by $L$, the order of approximation
), one calculates its critical temperature $T^c_L$ together with the
so-called critical mean-field coefficient $\bar Q_L$ for each
singular quantity $Q$.  The coefficient $\bar Q_L$ characterizes the
classical singular behavior of $Q$ in the vicinity of the critical
point:
\be
Q_L \sim \bar Q_L (T / T^c_L-1)^{\omega_{\rm class}}
\ee
where $\omega_{\rm class}$ stands for the mean-field value of the
corresponding critical index.  If the exact critical behavior is
characterized by the exponent $\omega$
\be
Q \sim (T / T^*-1)^{\omega}
\ee
with $T^*$ being the exact critical temperature, the mean-field
critical coefficients $\{\bar Q_L\}$ can be shown to obey the
following scaling formula \cite{Suzuki86, Suzuki87, Katori}
\be
\bar Q_L \sim (T^c_L - T^*)^{\omega-\omega_{\rm class}}
\ee
Thus, making use of the data $\{T^c_L,\bar Q_L\}$, we can obtain an
estimate for the exact critical index $\omega$.
For example, the values of the critical indices
$\alpha$, $\beta$, $\gamma$ and $\delta$
can be estimated from
\begin{eqnarray}
\bar c_L    & \sim & (T^c_L -T^*)^{-\alpha} \\
\bar m_L    & \sim & (T^c_L -T^*)^{\beta - 1/2} \\
\bar \chi_L & \sim & (T^c_L -T^*)^{1 - \gamma} \\
\bar m^c_L  & \sim & (T^c_L -T^*)^{-\psi} \; , \;\;\;
      \psi=\gamma(\delta-3)/3(\delta-1)
\end{eqnarray}
where $\bar c_L$, $\bar m_L$, $\bar \chi_L$ and $\bar m^c_L$ are
the mean-field critical coefficients of the specific heat, magnetization,
susceptibility and of the critical magnetization, respectively.
They fulfill the relations \cite{Suzuki86}
\be
\bar\chi_L\, \bar c_L /(\bar m_L)^2 = {\rm const}  \;\;\; {\rm and} \;\;\;
(\bar\chi_L)^2\, \bar c_L /(\bar m^c_L)^3 = {\rm const}
\label{const}\; ,
\ee
independently of the used series of approximations.
As a consequence, it is an inherent property of the CAM
that the scaling relations
$\alpha+2\beta+\gamma=2$ and $\gamma=\beta(\delta-1)$
are always satisfied by resulting exponents.

\subsection{Correction terms for the CAM data analysis}

In general, it is difficult to estimate the accuracy of the CAM
results.  This is mainly due to the fact that we do not know a
priori whether or not we are already in the asymptotic scaling region.
Usually, even very simple
mean-field type approximations show a quite good
coherent-anomaly, but one can always observe deviations from the
ideal CAM scaling.

{}From our experience with the CAM, we know that these deviations
depend very much on the specific series of approximations used
and are mainly of two kinds.  Sometimes one can detect a
deviation which changes smoothly with the
increasing order of the approximation.  In such a case it is
possible to weaken its effect by incorporating suitable corrections
into the CAM scaling formulas \cite{Katori}.
However, as a rule, the CAM data contain also departures from the
ideal scaling which seem to be erratic, without any clear
dependence on the order of the approximation.  Such a ``noise''
is usually negligible in comparison with other sources of errors
in our estimates, but as will be seen, it can be the limiting
factor in the accuracy of the present method even with a very
well behaved series of approximations.  Namely, if there are no
pronounced ``smooth'' deviations, the noise becomes discernible
and, because of its random character, it is hardly possible to improve
the results by introducing further corrections.

Before being limited by this subtle effect, however, one has to
tackle one more problem. Namely,
the results of the CAM analysis generally depend on
the choice of an independent variable. For instance,
\be
\bar\chi_L \sim \left({T^c_L-T^*}\over{T^*}\right)^{1-\gamma}
\;\;\; {\rm corresponds \ \ to} \;\;\;
\bar\chi_L \sim
\left({\beta^c_L}\over{\beta^*}\right)^{\gamma-1}
\left({\beta^*-\beta^c_L}\over{\beta^*}\right)^{1-\gamma}
\ee
in the inverse parameter $\beta=1/T$. The first factor on the right-hand
side produces a background contribution in the log-log
CAM plot when going from the temperature $T$ to the inverse temperature
$\beta$. This is why one obtains different estimates for critical
exponents using different variables.  This effect should vanish for
extremely good approximations, but this is not the case in practice.
Naturally, there is no distinguished ``right'' variable to be preferred.
Then, how to make our results insensitive to the choice of an
independent variable?  The idea is to introduce a correction factor which
cancels the background terms induced by transformation of variables:
\be
\bar\chi_L \sim
\left({x^c_L}\over{x^*}\right)^{\phi}
\left({\vert x^*-x^c_L\vert }\over{x^*}\right)^{1-\gamma} \; ,
\label{inv1}
\ee
where $x$ is a variable playing the role of the temperature or of its
inverse.  The exponent $\phi$ is determined from the requirement that
this formula is invariant under the exchange $x \leftrightarrow x^{-1}$.
It is easy to see that the appropriate value is given by
\be
\phi = (\gamma-1)/2
\label{inv2}
\ee
The critical exponent estimated in terms of the CAM formula
(\ref{inv1}) with (\ref{inv2}) is the same, whether we use $T$ or the
inverse temperature $\beta$.  Naturally, the following
formulas similar to the one for the susceptibility exponent can be used
for other critical indices:
\begin{eqnarray}
\bar c_L  & \sim &
\left({x^c_L}\over{x^*}\right)^{\alpha /2}
\left({\vert x^*-x^c_L\vert }\over{x^*}\right)^{-\alpha}
= \left(\vert \Delta_L \vert\right)^{-\alpha},
\label{cczac}
\\
\bar m_L  & \sim &
\left({x^c_L}\over{x^*}\right)^{(1/2-\beta)/2}
\left({\vert x^*-x^c_L\vert }\over{x^*}\right)^{\beta -1/2}
= \left(\vert \Delta_L \vert\right)^{\beta -1/2},
\\
\bar\chi_L  & \sim  &
\left({x^c_L}\over{x^*}\right)^{(\gamma-1)/2}
\left({\vert x^*-x^c_L\vert }\over{x^*}\right)^{1-\gamma}
= \left(\vert \Delta_L \vert\right)^{1-\gamma},
\\
\bar m^c_L  & \sim &
\left({x^c_L}\over{x^*}\right)^{\psi /2}
\left({\vert x^*-x^c_L\vert }\over{x^*}\right)^{-\psi}
= \left(\vert \Delta_L \vert\right)^{-\psi} ,\;
  \psi=\gamma (\delta-3)/3(\delta-1)
\label{cckon}
\end{eqnarray}
where $\Delta_L=(x^*/x^c_L)^{1/2}-(x^c_L/x^*)^{1/2}$.  Thus, we
propose to fit the critical exponents to the CAM data using the
variable $\Delta_L$ instead of
the usual temperature or its inverse.  It is worth to note that these
formulas are nearly invariant with respect to more
general transformation $x\rightarrow x^a$.  Namely, the change
$\Delta_L \to (x^*/x^c_L)^{a/2}-(x^c_L/x^*)^{a/2}$ does not produce
corrections linear in $x^c_L-x^*$ in the log-log CAM plot.  In the
present case, we have obtained practically the same
results with $x=\beta^{\pm a}$ for the values $a\in (0.1, 2)$ ; the
differences appear only at the sixth decimal place of the estimated
exponents, which is far beyond the accuracy of our estimates.  This
means that we can choose the variable $x$ in (\ref{cczac}) - (\ref{cckon})
quite arbitrarily without changing the estimates of critical indices.

{}From the practical point of view it is better to treat the exponent of the
correction term $\phi$ as a free parameter, and first to try to fit it together
with the usual exponent $\omega -\omega_{\rm class}$ (and possibly with $T^*$)
to the CAM data.  If the resulting parameters are stable with respect to
different subsets of the available data $\{T^c_L,\bar Q_L\}$, then it
means that some kind of the above-mentioned smooth background is observed.
In such a case, the fitted values of correction-term exponents should be
preferred to those listed in formulas (\ref{cczac}) - (\ref{cckon}).
However, if we cannot detect the background corrections reliably, we
have to resort to our invariant scheme (\ref{cczac}) - (\ref{cckon})
fixing the correction-term exponents.
Testing this scheme is one of the aims of the present work.

\section{Generation of mean-field approximations for CAM -- VSE method}

The crucial ingredient in the CAM is a series of approximations for,
let's say, the free energy of the system under investigation.  Such a
series is required to converge to its exact limit, and each
approximation should give a mean-field type solution of the model, i.e.
it should exhibit a classical singularity at its critical point.  Many
types of approximations have been used within the CAM approach, and it
seems that the best choice depends on the model studied.  In this work
we employ the so-called variational series expansion (VSE) method
\cite{Kolesik93a, Kolesik93b, Kolesik93c} for calculating our
approximations for the CAM.

\subsection{Mapping the Ising model onto a 256-vertex model}

Let us consider the simple-cubic Ising model described by the
Hamiltonian
\be
{\cal H} = -\sum_{<i,j>} s_i s_j - H \sum_i s_i
\ee
Instead of an external field $H$ we prefer to use the dimensionless field
$h=\beta H$ below.
The partition function
\be
Z = \sum_{\{s_i\}} \exp(-\beta \cal H)
\ee
can be rewritten as follows:
\be
\sum_{\{ s_{abc}\}} \prod_{(x y z)}^{*}
   w( s_{x y z}, s_{x+1 y z}, s_{x+1 y+1 z},  s_{x y+1 z},
      s_{x y z+1}, s_{x+1 y z+1}, s_{x+1 y+1 z+1},  s_{x y+1 z+1}  )
\ee
where the product runs only over the triples $(x y z)$ in which all
entries are either even or odd, and
\begin{eqnarray}
w(s_1,s_2,s_3,s_4,s_5,s_6,s_7,s_8)= & & \nonumber \\
\exp[ \beta (&s_1s_2+s_2s_3+s_3s_4+s_4s_1+& \nonumber \\
             &s_5s_6+s_6s_7+s_7s_8+s_8s_5+& \nonumber \\
             &s_1s_5+s_2s_6+s_3s_7+s_4s_8\phantom{+}&)]
\exp( h/2 \sum s_i )
\label{vahy}
\end{eqnarray}
is the contribution to the statistical weight coming from the eight
spins located in the corners of a cube.  These weights $\{w\}$ can be
considered
as the vertex weights of a 256-vertex model defined on a bcc lattice, in
which the original spins represent the edges of the bcc lattice.  The summation
over
all spin configurations is now understood as the summation over all
configurations of the edge states.

Vertex models are known to be gauge-invariant (see e.g \cite{Wegner}
\cite{Gaaf}).  Namely, their partition function remains unchanged when
one replaces the given vertex weights by the transformed ones.  We use
the following parameterization of the gauge transformation:
\be
\tilde w(s_1,\ldots,s_8) =
   \sum_{r_i=\pm 1} V_{s_1r_1}\ldots V_{s_8r_8} w(r_1,\ldots,r_8)
\ee
with $V$ being an orthogonal matrix dependent on the gauge parameter $y$:
\be
V(y) ={1\over \sqrt{1+y^2}}\pmatrix{1 & y \cr -y & 1\cr}
\label{vahyk}
\ee
Thus, the vertex weights can be considered as functions of the gauge parameter
$y$, and we do not distinguish between the original and the transformed weights
below.  It is the gauge parameter which is used for generating mean-field
approximations based on the formal series expansion for the vertex model.

Naturally, the sc Ising model can be transformed also onto a two-state
vertex model defined on the original lattice.  Our choice for the bcc lattice
has
the following motivation.  Firstly, we expect to obtain better
approximations with this representation.  For instance,
our lowest-order approximation (described below) is already better
than the Bethe approximation, and relatively small graphs embedded in
the coarse-grained lattice typically encompass many original spins.
Effectively, a single vertex in the bcc lattice represents four spins.
Secondly, in this way we arrive at a new classification scheme for
graphs contributing to our series expansions - very roughly speaking,
clusters are counted in an order different from that in the
formulation on the original lattice, and it turns out that the
resulting coherent anomaly is essentially free of pronounced
corrections to the CAM scaling.  From this point of view, we find that
the behavior of a series generated for the sc lattice is worse.

\subsection{Formal series expansion for the vertex model}

The second step in the VSE method is to calculate a formal series
expansion for the general 256-vertex model on the bcc lattice.
Thus, for the purpose of the generation of the expansion we need to
consider all the allowed weights as independent variables.  Because
of the symmetry of the lattice, the complete set of 256 vertex
weights $w(s_1,s_2,\ldots,s_8)$ consists of 22 equivalence classes
$\omega_i$ $(i=0,\ldots,21)$ represented e.g.  by
\begin{eqnarray}
 \omega_{ 0\phantom{0}}=w({\scriptstyle ++++++++ }) &
 \omega_{ 1\phantom{0}}=w({\scriptstyle +++-++++ }) &
 \omega_{ 2\phantom{0}}=w({\scriptstyle +-+-++++ })\nonumber \\
\phantom{\omega_{ 0}=}\vdots\phantom{w({\scriptstyle ++++++++ })} &
\phantom{\omega_{ 0}=}\vdots\phantom{w({\scriptstyle ++++++++ })} &
\phantom{\omega_{ 0}=}\vdots\phantom{w({\scriptstyle ++++++++ })}
\nonumber \\
 \omega_{18}=w({\scriptstyle -+---+-- }) &
 \omega_{19}=w({\scriptstyle ---+-+-- }) &
 \omega_{20}=w({\scriptstyle -----+-- })\nonumber \\
 \omega_{21}=w({\scriptstyle -------- }) &
 \phantom{\omega_{18}=w({\scriptstyle -+---+-- })} &
 \phantom{\omega_{19}=w({\scriptstyle ---+-+-- })}
\label{repre}
\end{eqnarray}

As a starting point for generating the expansion, we need a
ground-state.  At this stage, the convergence properties of the
formal expansion are irrelevant, and we can choose the configuration
with all edges in the state $+$ as our formal ground-state.  (Note
that this ground-state has nothing to do with the configuration with
all spin aligned up, because we have not specified the gauge yet.)
Each vertex in this configuration has the weight $\omega_0$ (see
(\ref{repre})), and, consequently, the ground-state contribution to
the dimensionless free energy per site is $-\log(\omega_0)$.

Next,
we have to take into account excitations above the ground-state.
They are represented by graphs weakly embedded in the lattice.  The
bonds connecting nodes of these graphs correspond to the edges in
the state $-$.  According to the states on its incident edges, each
vertex belonging to such a graph falls into one of the vertex
classes $\omega_i$, and the weight of the graph is given by $\prod_i
(\omega_i/\omega_0)^{n_i}$ where $n_i$ stands for the number of
those nodes which belong to the class $i$.

We use the no-free-ends
method described in Ref. \cite{Kolesik94} which allows us to eliminate the
so-called free-ends graphs from the calculation.  The free-ends
graphs are those which contain at least one node with a single
incident edge in the state $-$, i.e., the node of the class
$\omega_1$.  Thus, for calculating our series we have to generate
all possible no-free-ends graphs (up to a certain size $L$) embedded
in the bcc lattice.  Because the minimal number of nodes of a
no-free-ends graph is four, it is sufficient to consider only graphs
with two connected components for calculating the expansion up to
the order 11.  We calculated separately the contributions of graphs
with single and two connected components.  In this part of the
calculation we discriminated between the two sublattices in order to
make it possible to use the so-called code balance (i.e.  the
symmetry with respect to the sublattice exchange) for checking our
series.  We implemented a kind of a shadow method which treats the
two sublattices in a completely different way, such that the
sublattice symmetry is restored only in the final expansion, while
both the one- and two-component contributions lack this symmetry.
This turns to be a useful check to our calculations.

Having calculated the no-free-ends part of the series we have
generated the free-ends part by a simple algebraic procedure
explained in detail in Ref.  \cite{Kolesik94} .  After summing
all the contributions, the final series expansion has the form
\be
{\cal F}_L =  \log(\omega_0) + \sum_{n=2}^L
f_n\left(\left\{{ \omega_i \over\omega_0} \right\}_{i=1}^{21}\right)
\label{series}
\ee
Here (we use the logarithm of the statistical sum rather than the
free energy.), $f_n$ is a homogeneous polynomial of order $n$
representing the contributions of all graphs (connected or not)
with $n$ vertices, and $L$ denotes the maximal order included in
the expansion.  We have generated the expansion up to the order
$L=10$.  The final series consists of about $3\times10^4$ terms, and it
is therefore impossible to present it here \footnote{The series is
available upon request at the e-mail address
fyzikomi@shpa.phys.s.u-tokyo.ac.jp (till June 1995) and/or
fyzikomi@savba.savba.sk (after June 1995)}.

\subsection{From series expansions to mean-field approximations}

To extract physical information from the formal series expansion
(\ref{series}), we return to the weights (\ref{vahy}) -
(\ref{vahyk}) describing our original model.  They depend on the
inverse temperature $\beta$, external field $h$ and on the gauge
parameter $y$.  The dependence of the free energy on the gauge
should vanish in the limit $L\to \infty$, provided the limit
exists.  However, our truncated expansion ${\cal
F}_L(\beta,h,y)$ depends on $y$ for arbitrary finite order $L$.
In order to restore the gauge invariance of the free energy, at
least locally, we impose a  minimal-sensitivity condition
\be
{\partial {\cal F}_L(\beta,h,y) \over \partial y } = 0
\label{self}
\ee
This stationarity condition is a self-consistency equation for
determining the value of the gauge parameter within the VSE
scheme \cite{Kolesik93b}.  If there are more solutions to the
stationarity condition, we choose the one which
corresponds to the minimal free energy.

For the present model, it can be shown that $y=1$ is a solution
to (\ref{self}) for arbitrary temperature, provided the external
field vanishes.  In the low temperature region there exist two
more solutions which correspond to the ordered phase (these two
solutions differ in the sign of the magnetization).  The
critical point for a given order $L$ can be located easily as
the temperature at which this couple of low-temperature
solutions appear.  Analyzing the formal expansion ${\cal F}_L$
in the vicinity of the critical point, we can derive the
following formulas for the critical mean-field coefficients:
\begin{eqnarray}
\bar c_L &=&  \partial_{\beta\beta}{\cal F} -
            3 (\beta^c_L)^2 (\partial_{yy\beta}{\cal F})^2 /
            \partial_{yyyy}{\cal F}
            \label{cecko} \\
\bar m_L &=&  \partial_{hy}{\cal F}
             (-6 \beta^c_L \partial_{yy\beta} {\cal F}/
             \partial_{yyyy}{\cal F} )^{1/2}  \\
\bar \chi_L &=& (\partial_{hy}{\cal F})^2 /
             \partial_{yy\beta} {\cal F}\\
\bar m^c_L &=&  \partial_{hy}{\cal F}
             (-6 \beta^c_L \partial_{hy} {\cal F}/
             \partial_{yyyy}{\cal F} )^{1/3}
\end{eqnarray}
where all derivatives are taken at the critical point
$(\beta^c_L, h=0, y=1)$.

Among the critical coefficients, the specific-heat coefficient $\bar c_L$
deserves special attention.  The first term on right-hand side
of (\ref{cecko}) expresses the critical specific heat in the
high-temperature phase, and the complete formula corresponds to
the ordered phase.  In principle, the CAM scaling should be
observed in each of these terms.  Nevertheless, the
high-temperature term contains also the contribution from the
regular part of the specific heat.  This is why it does not
exhibit a good CAM scaling.  On the other hand, the second term
represents the difference between the high- and low-temperature
specific heats and is therefore free of the regular part.  For
this reason we restrict ourselves to the latter one.  Moreover,
in order to satisfy the relations (\ref{const}) strictly in each
order $L$, we have changed one of the $\beta^c_L$ factors into
exact critical value $\beta^*$.  Thus, we use
\be
\bar c_L = -3\beta^c_L\beta^* (\partial_{yy\beta}{\cal F})^2 /
             \partial_{yyyy}{\cal F}
\ee
instead of (\ref{cecko}).  As we shall see in the next Section,
we can arrive, in this way, at the estimate for the specific
heat exponent in very good agreement with the most precise results
available.  On the other hand we cannot extract any reliable
information from the specific heat in the disordered phase.

\section{CAM analysis}

We have calculated the mean-field critical coefficients (25) - (28)
for approximations with $L=0, \ldots , 10$.  The obtained values
together with the corresponding critical inverse temperatures
are listed in Table 1. The CAM plots are depicted in Fig.  1.
One can see that our series of approximation really exhibits a
very good coherent-anomaly.  Let us describe our analysis in
some detail.

For accurate estimation of the critical exponents within the CAM, one
needs a very precise value for the exact critical temperature.
Fortunately, for the Ising model we know the critical
temperature with high accuracy from MC studies.  In what follows
we use the value $\beta^*= 0.221652$ from the MCRG simulation
\cite{Baillie}.

First, we tried to find out whether or not there is some smooth
correction to the CAM scaling discernible in our data.  We
fitted our critical coefficients according to formulas
(\ref{cczac}) - (\ref{cckon}) without correction terms but with
added corrections to scaling as in Ref. \cite{Katori}.
Then we included the correction terms  but considered their exponents
as free parameters to be fitted.  In
both cases we were not able to fix the corrections to the CAM
scaling, because the resulting parameters as well as the
estimated exponents depended strongly on chosen subsets of
our CAM data; omitting just a single point often led to a
completely different output.  This means that the deviations of
the critical coefficients from the expected CAM scaling do not
show any systematic tendency but have a random character.
Consequently, we have to use our invariant scheme
(\ref{cczac}) - (\ref{cckon}).

Thus, we have fitted our critical coefficients, using formulas
(\ref{cczac}) - (\ref{cckon}), to various subsets of the available
data in order to compare the results.  The obtained estimates of the
critical exponents are listed in Table 2 together with specification
of the used data.

Having the critical coefficients without pronounced correction
to scaling, we tend to trust more the estimates based on the most
complete data set.  However, we check here for the stability
of the resulting critical exponents with respect to omitting some points.
The most reliable results are listed in the first part of Table 2.
One can see that the agreement between them is satisfactory.

Then we have calculated a list of estimates using only pairs of
points (see the second part of Table 2)  in order to get
feeling about the precision of our estimates.  Again, we see
that the agreement is quite good. (Naturally, exponents obtained
from near points, e.g.,  neighbors in the CAM plot, are often
rather different but there is no good reason to take such
``estimates'' seriously).

As was already mentioned, it is difficult to determine the
accuracy of the CAM estimates of exponents in general, because
one can never exclude a possibility of systematic deviations.
Nevertheless, we believe that the effect of possible corrections
to scaling causing systematic deviations is buried in the
noise of the CAM data in the present case.  Consequently,
the comparison between the results obtained from different
subsets of the available data could give a good estimate
for errors of our critical exponents.  In this way we arrive at
our final estimates as
\be
\alpha=\afin,\;\;\; \beta=\bfin,\;\;\;
\gamma=\gfin,\;\;\; \delta=\dfin,\;\;\; \psi=\pfin ,
\label{real}
\ee
where the values come from our most reliable fit (1-7) (see Table
2) and the error bars are chosen such that they essentially cover
all the most significant estimates in Table 2.  Alternatively, we have
calculated mean values from the estimates listed in the second part of
Table 2 and obtained the results
\be
\alpha=\aopt,\; \beta=\bopt,\;
\gamma=\gopt,\; \delta=\dopt,\; \psi=\popt
\label{opt}
\ee
where errors quoted correspond just to the standard deviations.
It should be remarked that the estimated values (\ref{opt}) have
smaller error bars but taking mean values is not necessarily
justified.  We, therefore, prefer the values (\ref{real}).
We would like to stress again that our results are consistent
with the scaling relations
$\alpha+2\beta+\gamma=2$ and $\gamma=\beta(\delta-1)$.
The values of the critical exponents $\nu$ and $\eta$ calculated
from (\ref{real})  and the scaling relations
$d\nu=2-\alpha$, $\eta=2-\gamma/\nu$
are $\nu=0.631$ and $\eta=0.039$.

Finally, we present some recent estimates of 3D Ising critical
exponents in Table 3 for comparison with our results.  We can
conclude that our present estimates are in agreement with
results of other methods within the error bars and that our
accuracy is comparable to the one of the most accurate previous
results. In comparison with other CAM-based
works, on the other hand, the present results show distinct improvement.

\section{Concluding remarks}

In the present work we have shown that the coherent-anomaly
method can compete with the most accurate methods based on
series expansions and/or large-scale computer simulations.  It
turns out that the choice of the series of classical
approximations used in the CAM analysis is very important.  The
present improvement in comparison with the previous CAM-based
studies was achieved mainly owing to the new type of the series
expansion used here, which exhibits a coherent-anomaly essentially
free of systematic corrections.

The second important point in the present approach is the
use of the invariant scheme (\ref{cczac}) - (\ref{cckon}).
It is useful
whenever we have the mean-field critical coefficients for which
it is not possible to extract the corrections to the CAM scaling
reliably.  In such a case it represents the most natural way for
resolving the ambiguity related to the choice of the independent
variable.
Without using the correction terms we would obtain, for
instance, the values $\gamma=1.2395(25)$ and $\gamma=1.2350(23)$
(compare them to (\ref{opt})) from  fitting in the inverse temperature
$\beta$ and in the temperature $T$, respectively.  Similarly,
also other indices would be systematically different, depending on
the variable used.

The bottleneck of the whole calculation is the generation of the
series expansion.  We calculated it up to the order $L=10$ using
only about three hours of CPU time of a small HP workstation.
Thus, the computer demands of the present method are essentially
smaller than in the methods based on series expansions as well
as on MC simulations.  This immediately rises the question
whether it is possible to improve our results by using a faster
machine, more computer time and more memory space in order to extend
our series.  The answer is, unfortunately, no.
That is because the uncertainty in our estimates is dominated by
the chaotic deviations from the ideal CAM scaling, and we would
need many new terms to make this effect smaller.  We have
estimated that  we would need more than
20 days with the same computer to extend the series up to order 13,
but that would not be
sufficient to reduce error bars significantly.  From
this follows that  we had better
try to reduce the noise in the CAM data rather than to use brute force.

One possible way could be perhaps to extend  the variational
series expansion scheme in the sense of correlated mean-field
theories, or towards a (quasi)continuous family of classical
approximations, similarly as in Ref. \cite{Cenedese}.  We hope to
report on progress in this direction in the future.

\newpage

\newlength{\del}
\setlength{\del}{0.05cm}
\newlength{\dlz}
\setlength{\dlz}{\baselineskip}
\addtolength{\dlz}{\del}

\centerline{\large\bf Table 1.}
\bigskip
\begin{center}
\begin{tabular}{|c|c|c|c|c|c|c| } \hline
\rule[-0.25cm]{0cm}{0.7cm} No.&$L$&$\beta^c_L$&$\bar c_L$&
$\bar m_L$&$\bar \chi_L$&$\bar m^c_L$  \\ \hline
\hline
\rule{0cm}{\dlz}0 & 0-3 & 0.206633 & 6.848577 & 8.181219 &
 1.083122 & 4.169696 \\ 
1 & 4 & 0.214738 & 7.599206 & 9.563006 & 1.333713 & 4.959262 \\ 
2 & 5 & 0.214844 & 7.650375 & 9.622822 & 1.341417 & 4.989491 \\ 
3 & 6 & 0.216791 & 7.944006 & 10.24680 & 1.464803 & 5.357788 \\ 
4 & 8 & 0.217938 & 8.251377 & 10.74478 & 1.550638 & 5.635969 \\ 
5 & 7 & 0.218671 & 8.316348 & 11.07017 & 1.633118 & 5.849366 \\ 
6 & 9 & 0.219200 & 8.521609 & 11.47035 & 1.711093 & 6.083345 \\ 
7\rule[-0.3cm]{0pt}{5pt} & 10& 0.219384 & 8.643945 & 11.67254 &
 1.746871 & 6.197225 \\ \hline
\end{tabular}
\end{center}
\vfill

{\bf Table 1.}  Critical mean-field coefficients for the specific heat, $\bar
c_L$, magnetization, $\bar m_L$, susceptibility, $\bar\chi_L$, and for the
critical
magnetization, $\bar m^c_L$.  (Approximations of the order $L=0,1,2,3$
have the same classical critical behavior within the VSE method and were
excluded from the analysis from the very beginning.)

\newpage

\centerline{\large\bf Table 2.}
\bigskip
\begin{center}
\begin{tabular}{|c|c|c|c|c| }
\multicolumn{4}{l}{most reliable estimates}\\ \hline
\rule[-0.25cm]{0cm}{0.7cm} data&$\alpha$&$\beta$&\ $\gamma$\ &$-\psi$ \\ \hline
\hline
\rule{0cm}{\dlz}1 -- 4  & 0.12598 & 0.31645 & 1.24111 & 0.20273 \\ 
1 -- 5  & 0.10991 & 0.32650 & 1.23709 & 0.19470 \\ 
1 -- 6  & 0.10721 & 0.32839 & 1.23602 & 0.19308 \\ 
1 -- 7  & 0.10783 & 0.32736 & 1.23655 & 0.19394 \\ 
2 -- 4  & 0.12279 & 0.31996 & 1.23728 & 0.19912 \\ 
2 -- 5  & 0.10514 & 0.33050 & 1.23386 & 0.19095 \\ 
2 -- 6  & 0.10321 & 0.33167 & 1.23345 & 0.19004 \\ 
2 -- 7\rule[-0.3cm]{0pt}{5pt}& 0.10563 & 0.32994 & 1.23448 & 0.19153 \\ \hline
\multicolumn{5}{l}{two-point estimates }\\ \hline
\rule[-0.25cm]{0cm}{0.7cm} data&$\alpha$&$\beta$&$\gamma$&$-\psi$ \\ \hline
\hline
\rule{0cm}{\dlz}1 , 3  & 0.12424 & 0.30662 & 1.26251 & 0.21642 \\ 
1 , 4  & 0.13092 & 0.31473 & 1.23962 & 0.20339 \\ 
1 , 5  & 0.10604 & 0.32791 & 1.23814 & 0.19411 \\ 
1 , 6  & 0.10942 & 0.32630 & 1.23798 & 0.19513 \\ 
1 , 7  & 0.11443 & 0.32292 & 1.23973 & 0.19796 \\ 
2 , 4  & 0.12334 & 0.32014 & 1.23638 & 0.19870 \\ 
2 , 5  & 0.10000 & 0.33213 & 1.23574 & 0.19049 \\ 
2 , 6  & 0.10458 & 0.32969 & 1.23603 & 0.19222 \\ 
2 , 7  & 0.11001 & 0.32602 & 1.23794 & 0.19530 \\ 
3 , 5  & 0.09286 & 0.34332 & 1.22050 & 0.17795 \\ 
3 , 6  & 0.10174 & 0.33649 & 1.22528 & 0.18410 \\ 
3 , 7  & 0.10987 & 0.33049 & 1.22914 & 0.18939 \\ 
4 , 6  & 0.07708 & 0.34370 & 1.23552 & 0.18270 \\ 
4 , 7  & 0.09356 & 0.33329 & 1.23987 & 0.19110 \\ 
5 , 7\rule[-0.3cm]{0pt}{5pt}& 0.14037 & 0.30750 & 1.24464 & 0.20988 \\ \hline
\end{tabular}
\end{center}
\vfill

{\bf Table 2.}  \nopagebreak In the first part of this table we have
listed our most reliable estimates for critical exponents using
different subsets of our CAM-data.  The notation $x - y$ in the first
column means that all the points from $x$ up to $y$ (numbered as in
Table 1) were used for calculating the exponents.  In the second part of
Table 2 we have listed estimates as obtained using only two points
indicated in the first column.

\newpage

\centerline{\large\bf Table 3.}
\bigskip
\begin{center}
\begin{tabular}{|l|c|l|l|l|l|l|} \hline
\rule[-0.25cm]{0cm}{0.7cm} Author & year & Ref. &\hfil $\alpha$ &\hfil
                 $\beta$ &\hfil  $\gamma$ &\hfil  $\delta$ \\ \hline
Adler & 1983 & \cite{Adler83} &\rule{0cm}{\dlz} & & 1.239(3) & \\
Adler & 1982 & \cite{Adler82} & & & 1.238(3) & \\
Bhanot & 1994 & \cite{Bhanot94} & 0.104(4) & & & \\
Bhanot & 1992 & \cite{Bhanot92} & 0.207(4) & 0.308(5) & & \\
Ferenberg & 1991 & \cite{Ferenberg} & & 0.3258 & 1.239(7) & \\
Guttmann & 1994 & \cite{Guttmann94} & 0.101(4) &  & & \\
Guttmann & 1993 & \cite{Guttmann93} & 0.110(5) & 0.329(9) & & \\
Guttmann & 1987 & \cite{Guttmann87,Guttmann87a} & 0.104(6) & &
   1.239(3) & \\
Guttmann & 1986 & \cite{Guttmann86} & & & 1.240(6) & \\
Ito & 1991 & \cite{Ito} & & 0.324(4) & & \\
Le Guillou & 1987 & \cite{LeGuillou87} & & 0.3270(15) & 1.2390(25) & \\
Le Guillou & 1980 & \cite{LeGuillou80} & & 0.3250(15) & 1.2341(20) & \\
Nickel & 1991 & \cite{Nickel91} & & & 1.238 & \\
Nickel & 1990 & \cite{Nickel90} & 0.11(2) & & 1.237(2) & \\
Oitmaa & 1991 & \cite{Oitmaa} & 0.096(6) & 0.318(4) & 1.25(2) & \\
Ruge & 1994 & \cite{Ruge} &\rule[-0.3cm]{0pt}{5pt} & 0.319(5) &
    1.237 & \\ \hline
Cenedese & 1994 & \cite{Cenedese}$^*$\rule{0cm}{\dlz}
     & 0.105 & 0.318 & 1.258 & 4.957 \\
     &&& 0.075 & 0.297 & 1.329 & 5.471 \\
Fujiki & 1990 & \cite{Fujiki}$^*$ & & & 1.276 & \\
Katori & 1994 & \cite{Katori94}$^*$ & & & 1.296 & \\
Katori & 1987 & \cite{Katori}$^*$ & & & 1.258 & \\
Kinosita & 1992 & \cite{Kinosita}$^*$ & & & 1.246 & \\
Kolesik & 1993 & \cite{Kolesik93c}$^*$ & & 0.331 & 1.238 &  \\
Monroe & 1988 & \cite{Monroe}$^*$ & & 0.35 & 1.32 & \\ \hline
this work &\rule{0cm}{\dlz} &\rule[-0.3cm]{0pt}{5pt}$^*$
          & \afin & \bfin & \gfin & \dfin\\ \hline
\end{tabular}
\end{center}
\vfill

{\bf Table 3.}  Some recent estimates for critical exponents of the
3D Ising model.  We have included also works based on the CAM
(denoted by $^*$) for comparison.  Our final estimates are listed
in the last row of the table.

\newpage
\section{Figure caption}

{\bf Fig.  1.}  CAM scaling of the critical mean-field coefficients
for the specific heat, $\bar c_L$ (+), magnetization, $\bar m_L$
($\Box$), susceptibility, $\bar \chi_L$ ($\Diamond$) and for the critical
magnetization, $\bar m^c_L$ ($\times$).  The distance from the true
critical point is measured in $\Delta_L =
(\beta^*/\beta^c_L)^{1/2}-(\beta^c_L/\beta^*)^{1/2}$.  Critical
coefficients were rescaled in order to get them into the same
plotting area.  The straight lines correspond to our most reliable fit
(1-7) (see Table 2).

\vfill
\newpage

\newcommand{\rf}[5]{{\rm #1, }{ #2 }{ #3} (#4) #5. }


\begin{thebibliography}{99}

\bibitem{Suzuki86}
        \rf{M. Suzuki}
           {J. Phys. Soc. Jpn.}{55}{1986}{4205}
\bibitem{Suzuki87}
        \rf{M. Suzuki, M. Katori and X. Hu}
           {J. Phys. Soc. Jpn.}{56}{1987}{3092}

\bibitem{Katori}
        \rf{M. Katori and M. Suzuki}
           {J. Phys. Soc. Jpn.}{56}{1987}{3113}

\bibitem{Suzuki94}
        \rf{M. Suzuki, K. Minami and Y. Nonomura}
           {Physica A}{205}{1994}{80}

\bibitem{Kolesik93a}
        \rf{M. Kolesik and L. \v Samaj}
           {J. Phys. I (France)}{3}{1993}{93}
\bibitem{Kolesik93b}
        \rf{M. Kolesik and L. \v Samaj}
           {J. Stat. Phys.}{72}{1993}{1203}
\bibitem{Kolesik93c}
        \rf{M. Kolesik and L. \v Samaj}
           {Phys. Lett. A}{177}{1993}{87}

\bibitem{Wegner}
        \rf{F. J. Wegner}
           {Physica }{68}{1973}{570}

\bibitem{Gaaf}
        \rf{A. Gaaf and J. Hijmans }
           {Physica A}{80}{1975}{149}

\bibitem{Kolesik94}
        \rf{M. Kolesik}
           {Physica A}{202}{1994}{529}

\bibitem{Baillie}
        \rf{C. F. Baillie, R. Gupta, K. A. Hawick and G. S. Pawley}
           {Phys. Rev. B}{45}{1992}{10438}

\bibitem{Adler83}
        \rf{J. Adler}
           {J. Phys. A}{16}{1983}{3585}
\bibitem{Adler82}
        \rf{J. Adler, Moshe M and Privman V}
           {Phys. Rev. B}{26}{1982}{3958}

\bibitem{Bhanot94}
        \rf{G. Bhanot, M. Creutz, U. Gl\" assner and K. Schilling}
           {Phys. Rev. B}{49}{1994}{12909}
\bibitem{Bhanot92}
        \rf{G. Bhanot, M. Creutz and J. Lacki}
           {Phys. Rev. Lett.}{69}{1992}{1841}

\bibitem{Ferenberg}
        \rf{A. M. Ferrenberg and D. P. Landau }
           {Phys. Rev. B}{44}{1991}{5081}

\bibitem{Guttmann94}
        \rf{A. J. Guttmann and I. G. Enting}{}{}{COND-MAT/9411002}{}
\bibitem{Guttmann93}
        \rf{A. J. Guttmann and I. G. Enting}
           {J. Phys. A}{26}{1993}{807}
\bibitem{Guttmann87}
        \rf{A. J. Guttmann}
           {J. Phys. A}{20}{1987}{1855}
\bibitem{Guttmann87a}
        \rf{A. J. Guttmann}
           {J. Phys. A}{20}{1987}{1839}
\bibitem{Guttmann86}
        \rf{A. J. Guttmann}
           {Phys. Rev. B}{33}{1986}{5089}

\bibitem{Ito}
        \rf{N. Ito and M. Suzuki}
           {J. Phys. Soc. Jpn.}{60}{1991}{1978}

\bibitem{LeGuillou87}
        \rf{J. C. Le Guillou and J. Zinn-Justin }
           {J. Phys. (Paris)}{48}{1987}{19}
\bibitem{LeGuillou80}
        \rf{J. C. Le Guillou and J. Zinn-Justin }
           {Phys. Rev. B}{21}{1980}{3976}

\bibitem{Nickel90}
        \rf{B. G. Nickel and J. J. Rehr }
           {J. Stat. Phys.}{61}{1990}{1}
\bibitem{Nickel91}
        \rf{B. G. Nickel}
           {Physica}{177}{1991}{189}

\bibitem{Oitmaa}
        \rf{J. Oitmaa, C. J. Hamer and W. Zheng }
           {J. Phys. A}{24}{1991}{2863}

\bibitem{Ruge}
        \rf{C. Ruge, P. Zhu and F. Wagner }
           {Physica A}{209}{1994}{431}

\bibitem{Cenedese}
         \rf{P. Cenedese, J. M. Sanchez and R. Kikuchi }
            {Physica A}{209}{1994}{257}

\bibitem{Fujiki}
        \rf{S. Fujiki, M. Katori and M. Suzuki}
           {J. Phys. Soc. Jpn.}{59}{1990}{2681}

\bibitem{Katori94}
        \rf{M. Katori and M. Suzuki}
        {Prog. Theor. Phys. Suppl.}{115}{1994}{83}

\bibitem{Kinosita}
        \rf{Y. Kinosita, N. Kawashima  and M. Suzuki}
           {J. Phys. Soc. Jpn.}{61}{1992}{3887}

\bibitem{Monroe}
        \rf{J. L. Monroe}
           {Phys. Lett. A}{131}{1988}{427}

\end{thebibliography}
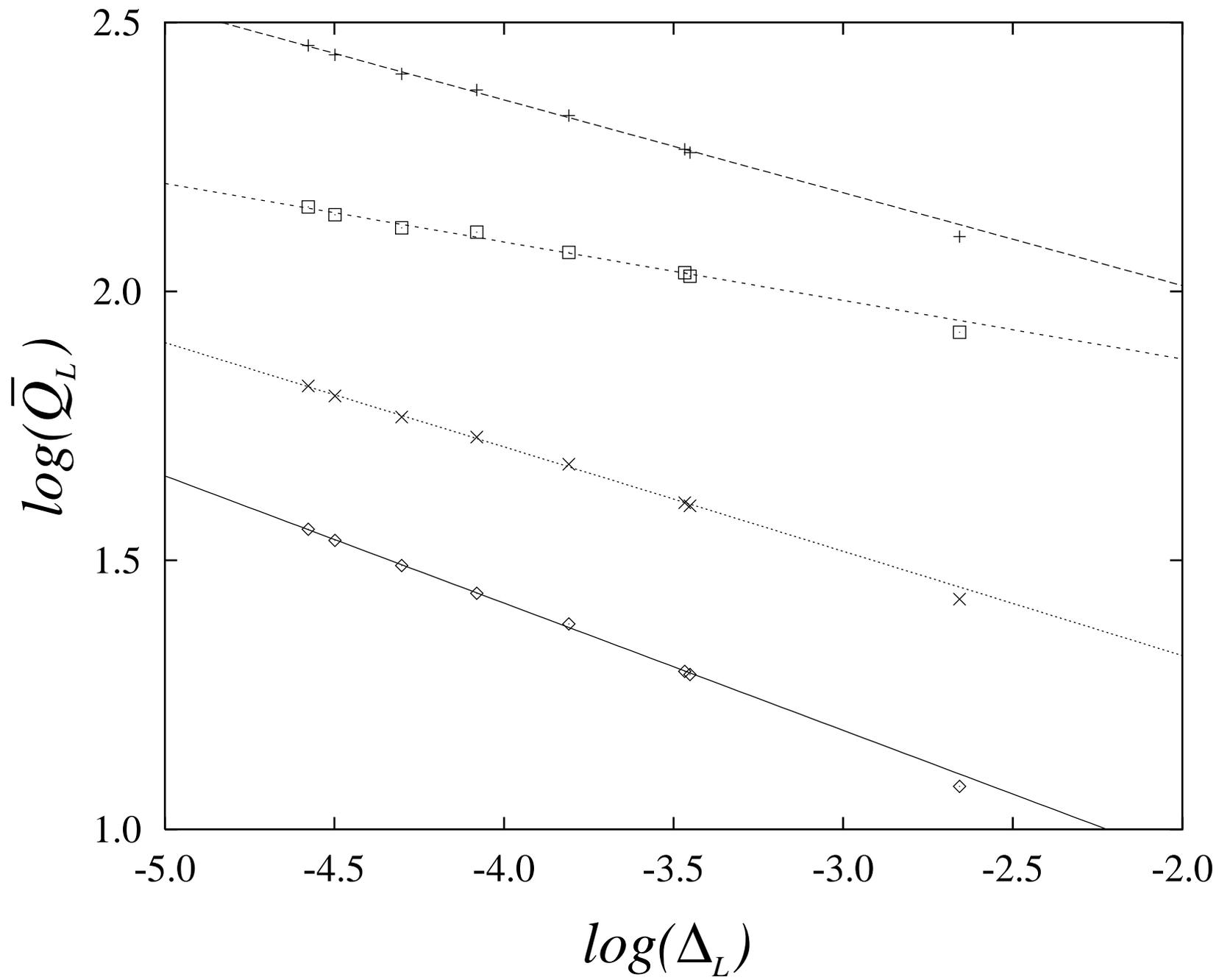
\end{document}